# How does COVID-19 change insurance and vaccine demand? Evidence from short-panel data in Japan

Short title: COVID-19, insurance, and vaccine demand.

Eiji Yamamura[¶1*], Yoshiro Tsutsumi[2]

[1]*Seinan Gakuin University, (Department of Economics), Fukuoka, (6-2-92 Sawaraku Nishijin), 814-8511, Japan*

[2]*Kyoto-Bunkyo University, Japan.*

*Corresponding author

Email: yamaei@seinan-gu.ac.jp

# Abstract

In this study, we explored how the coronavirus disease (COVID-19) affected the demand for insurance and vaccines in Japan from mid-March to mid-April 2020. Through independent internet surveys, respondents were asked hypothetical questions concerning the demand for insurance and vaccines for protection against COVID-19. Using the collected short-panel data, after controlling for individual characteristics using the fixed effects model, the key findings, within the context of the pandemic, were as follows: (1) Contrary to extant studies, the demand for insurance by females was smaller than that by their male counterparts; (2) The gap in demand for insurance between genders increased as the pandemic prevailed; (3) The demand for a vaccine by females was higher than that for males; and (4) As COVID-19 spread throughout Japan, demand for insurance decreased, whereas the demand for a vaccine increased.

# Introduction

Since the end of 2019, society has suffered from the tremendous global impact of the COVID-19 pandemic. Within six months, the spread of the disease changed lifestyles and economic activities (especially in the UK, Italy, Spain, France, the US, and Asian countries such as Japan, China, and Korea) as the number of COVID-19 victims drastically increased.

Existing research [1-5] has found that events such as natural disasters change attitudes and perceptions toward risk. Other studies [6-8] explored the impacts of conflicts on risk attitude. Yamamura [9,10] found that nuclear accidents changed risk perceptions about nuclear plants while increasing the demand for insurance [11]. Unlike natural disasters, which are usually confined to smaller geographical areas, the impact of this pandemic on society is predicted to differ in its extent. Extant studies have provided evidence that females are more likely to be risk-averse [12,13]. It is valuable to investigate whether the impact of unexpected events, such as COVID-19, affects risk attitude towards insurance differently between genders.

Consequently, the following questions arise: First, how does COVID-19 change the risk attitude of individuals, and second, how does the influence of COVID-19 differ between genders? No previous research has dealt with these questions, although some studies have analyzed the impact of COVID-19 on the mental health and perception of individuals [14-19].

Generally, more risk-averse individuals are predicted to demand non-life insurance. The COVID-19 pandemic creates a unique environment where the perceived value of money might be eroded when people believe their life is in jeopardy with an increased probability of death. Those who are more risk-averse may be less likely to demand non-life insurance under these circumstances, while life security becomes a higher priority in times of peril. In exploring the above inference, this study examines the impact of COVID-19 on demand for non-life insurance and vaccines using purposefully collected individual-level data. The panel data was constructed from three rounds of internet surveys conducted between March 13 and April 10, and even within this limited time frame, we could detect changes in non-life insurance and the demand for a vaccine.

Our key findings are as follows: First, the demand by females for non-life insurance was smaller than that by their male counterparts. Second, their demand for a vaccine against COVID-19 was higher than that of men. As COVID-19 prevailed throughout Japan, non-life insurance demand decreased, whereas vaccine demand increased.

## Materials and methods

### Design of surveys

The research company INTAGE was commissioned to conduct an internet survey for this study based on their experience and reliability. The sampling method was designed to gather a representative sample of the Japanese population in terms of gender, age, educational background, and residential area. Japanese citizens aged 16–79 years from across Japan were selected for the survey. To construct the panel data, online surveys were conducted at three separate times ("waves") within one month, targeting the same individuals.

The first wave of queries was conducted from March 13-16, recording 4,359 observations with a response rate of 54.7 %. The second wave was conducted from March 27-30, and the third on April 10-13. Respondents from the first wave were targeted in the second and third waves to track and record any changes in their attitudes over time. The second and third waves achieved response rates of 80.2% and 92.2%, respectively, from a total of 11,867 observations.

Fig 1 illustrates the change in the total number of persons infected with COVID-19 in Japan during the period from March 1 to April 20, indicating a modest increase between the first and the second waves, with a significant increase afterward. The

Japanese Government declared a state of emergency on April 10 to cope with the surge of infected persons. The declaration requested people to avoid leaving their homes unnecessarily and effected the closure of public places, including schools, museums, theaters, and bars, with the intention for this to be in effect for one month.

**Fig 1. Changes in total infected persons in Japan between March 1 and April 18.**
The first, second and third waves were conducted on Mach 10, March 27, and April 10, respectively. The State of Emergency Declaration for COVID-19 was promulgated on April 10, 2020.
Source: The total number of persons infected by COVID 19 is sourced from Johns Hopkins University of Medicine, CORONAVIRUS RESOURCE CENTER. https://coronavirus.jhu.edu/map.html. (On May 9, 2020).

## Measurements

Table 1 contains a description of the variables used in this paper. We asked respondents their demand for "hypothetical non-life insurance" and "hypothetical vaccines." Two "hypothetical non-life insurance" options were given: one for high probability and small associated loss (*Normal Insurance Demand*) and another for low probability and large associated loss (*Mega Insurance Demand*). According to basic economic theory, we expect demand to increase with an increase in these values because respondents will be prepared to pay more for the product.

**Table 1. Definitions of key variables and their basic statistics**

| Variables | Definition | Mean Male | Mean Female |
|---|---|---|---|
| *Normal Insurance Demand* | Assume that you know there is a 50% chance of losing 0.1 million yen ($ 1000) on a given day. You can take out insurance to cover this amount in case of a loss. What is the most you would pay to purchase the insurance? Choose from 11 choices.<br>1 (0 yen) - 11 (over 50 thousand yen).<br>In order to standardize the price respondents would pay, the price is divided by "losing money" as follows:<br>Prices chosen/0.1 million | 0.16 | 0.13 |
| *Mega Insurance Demand* | Assume that you know there is a 0.1% chance of losing 5 million yen ($ 0.5 million) on a given day. You can take out insurance to cover this amount in case of a loss. What is the most you would pay to purchase the insurance? Choose from 11 choices.<br>1 (0 yen) - 11 (over one million yen).<br>To standardize the price respondents would pay, the price is divided by "losing money" as follows: Prices chosen/5 million | 4.34 | 3.05 |
| *Vaccine demand* | Assume that a vaccine, which is verified to be effective against novel coronavirus, is available at the cost of 100 thousand yen (the cost is not covered by insurance). Would you take it after paying 100 thousand yen? Choose from 5 choices.<br>1 (definitely do not take) to 5 (definitely take) | 2.24 | 2.34 |

Figs 2a, 2b, and 3 illustrate the change in non-life insurance and vaccine demand after time has passed and compares the difference between genders. Fig 2 indicates that *Normal Insurance Demand* by females is lower than that of males in waves 1-3, which is in agreement with existing studies [12, 13]. Further, *Normal Insurance Demand* by females decreased more rapidly than men as COVID-19 spread. In the first wave, there is no statistical difference between genders; however, this changes in Waves 2 and 3. The same trend is also observed in the case of *Mega Insurance Demand.*

**Fig 2a and Fig 2b. Change of insurance demand as COVID-19 spread by type of loss.** The interval represents the 95% confidence intervals.
**Fig 3. Change of vaccine demand as COVID-19 spreads.** The interval represents the 95%

confidence intervals.

Fig 3 demonstrates *Vaccine Demand* among females is larger than for men in waves 1-3. Further, *Vaccine Demand* increases for both males and females as COVID-19 spreads.

## Methods

We used fixed effects model regression to control the time-invariant individual fixed effects. The estimated function takes the following form:

$$Y_{itg} = \alpha_1 Wave2_t + \alpha_2 Wave3_t + \alpha_3 \text{ Infected COVID-19}_{itg} + k_i + u_{itg},$$

In this formula, $Y_{itp}$ represents the dependent variable for individual *i*, wave *t*, and group *g*. *Y* is *Insurance demand* and *Vaccine demand*. The second (*Wave2*) and third (*Wave3*) wave dummies were included, with the first wave as the reference group. This approach captures the degree of change in the dependent variables compared to the first wave. The number of persons infected with COVID-19 was also included as a control variable.

The time-invariant individual-level fixed effects are represented by $k_i$. As short-term panel data was used, most of the individual-level demographic variables were considered as time-invariant features, which were completely described by $k_i$. The regression parameters are denoted as *α*. The error term is denoted as *u*.

# Results

## Non-life insurance demand

Table 2 reflects that both *Wave2* and *Wave3* have mostly negative values for their coefficients and statistical significance. In estimations of *Normal Insurance Demand* using the full sample, the absolute value of the coefficients of *Wave3* is 0.57, almost three times larger than those of *Wave2*. This implies that non-life insurance demand decreases at an accelerated pace. This also agrees with results where the sample is divided into male and female sub-samples, as indicated in columns 2 and 3. The absolute values of the coefficients based on the female sample are larger than those of the male sample. Similar results are obtained in the estimations of the *Mega Insurance Demand*.

**Table 2. Fixed effects model. Normal and Mega Insurance Demand.**

|  | Normal | | | Mega Risk | | |
|---|---|---|---|---|---|---|
|  | *Full* | *Male* | *Female* | *Full* | *Male* | *Female* |
| *Wave1* |  | < default > |  |  | < default > |  |
| *Wave2* | −0.018*** | −0.015** | −0.020*** | −0.926** | −0.148 | −1.692*** |
|  | (0.004) | (0.006) | (0.005) | (0.442) | (0.727) | (0.508) |
| *Wave3* | −0.057*** | −0.052*** | −0.062*** | −2.469*** | −1.945*** | −2.986*** |
|  | (0.004) | (0.007) | (0.005) | (0.445) | (0.701) | (0.551) |
| *Infected COVID19* | 0.006 | 0.011 | 0.001 | 0.279 | 0.932 | 1.457 |
|  | (0.008) | (0.013) | (0.011) | (0.861) | (1.585) | (0.703) |
| R-Square | 0.03 | 0.02 | 0.05 | 0.01 | 0.01 | 0.01 |
| Groups | 4,359 | 2,154 | 2,205 | 4,359 | 2,154 | 2,205 |
| Obs. | 11,867 | 5,880 | 5,987 | 11,867 | 5,880 | 5,987 |

Note: Numbers within parentheses are robust standard errors clustered on individuals. *** and ** indicate the statistical significance at 1% and 5% levels, respectively. R-square is "Within R-square."

# Vaccine demand

Table 3 indicates that both *Wave2* and *Wave3* show positive signs of their coefficient and statistical significance. Furthermore, in the results using the full sample, the absolute value of the coefficients of *Wave3* is 0.43, almost 1.8 times larger than those of *Wave2*. The absolute values of the coefficients of *Wave2* and *Wave3,* based on female samples, are 0.25 and 0.44, respectively, slightly larger than those based on male samples. This also holds for results when the sample is divided into male and female sub-samples, as presented in columns 2 and 3. Consistent with Fig 3, vaccine demand increased as COVID-19 spread, with a higher vaccine demand amongst females than males. Jointly analyzing Tables 2 and 3 indicates that people would reallocate their expenditure from non-life insurance to a vaccine as people felt their lives were in danger. Further, compared to males, females put more importance on their lives over their money.

**Table 3. Fixed effects model. Vaccine Demand.**

|  | Vaccine | | |
|---|---|---|---|
|  | *Full* | *Male* | *Female* |
| *Wave1* |  | < default > |  |
| *Wave2* | 0.246*** | 0.245*** | 0.247*** |
|  | (0.017) | (0.025) | (0.023) |
| *Wave3* | 0.425*** | 0.413*** | 0.438*** |
|  | (0.020) | (0.029) | (0.027) |
| *Infected COVID-19* | 0.067 | 0.058 | 0.075 |
|  | (0.041) | (0.058) | (0.057) |
| Within R-Square | 0.09 | 0.08 | 0.10 |
| Groups | 4,359 | 2,154 | 2,205 |
| Obs. | 11,867 | 5,880 | 5,987 |

Note: Number within parentheses is robust standard errors clustered on individuals. ***, ** and * indicates the statistical significance at 1%, 5% and 10 % levels, respectively.

## Discussion

Individuals place more importance on their lives than their money. Individuals who are risk-averse are more likely to avoid loss of life than loss of money as the probability of death rises. Under the coronavirus pandemic, individuals face risk to their lives, which is far from normal conditions. This is considered to be the reason that demand for non-life insurance by females was smaller than demand by males, and increased as COVID-19 spread. Conversely, the demand for a vaccine against COVID-19 by females was larger than for males, and this gap increased as COVID-19 spread.

As widely acknowledged, females are more risk-averse than males. Considering that people are in mortal danger, females, when compared to males, more readily allocate their money to the vaccine than to non-life insurance.

## Acknowledgments

We would like to thank Editage [http://www.editage.com] for editing and reviewing this manuscript for English language.